%% file: kronfeld.tex
\documentstyle[ask,epsfig,color]{aipproc}

\input MyBWMap.ltx

\def\mr{\mathrm}

\def\vek#1{\mbox{\protect\boldmath $#1$}}
\def\LQCD{\Lambda_{\mr QCD}}

\def\lsim{\mathop{\raisebox{-.4ex}{\rlap{$\sim$}} \raisebox{.4ex}{$<$}}}

\begin{document}
\begin{flushright}
FERMILAB-CONF-98/393-T
\end{flushright}
\title{Lattice QCD Calculations of Leptonic and Semileptonic Decays}

\author{Andreas S.~Kronfeld}
\address{Fermi National Accelerator Laboratory,%
\thanks{Fermilab is operated by Universities Research
Association Inc., under contract with the United States' Department of
Energy.}
Batavia, Illinois, U.S.A.%
}

\maketitle

\begin{abstract}
In lattice QCD, obtaining properties of heavy-light mesons has been
easier said than done.
Focusing on the $B$ meson's decay constant, it is argued that towards
the end of 1997 the last obstacles were removed, at least in the
quenched approximation.
These developments, which resulted from a fuller understanding and
implementation of ideas in effective field theory, bode well for current
studies of neutral meson mixing and of semileptonic decays.
\end{abstract}

\pagestyle{plain}

\section*{Introduction}
Eleven years ago at {\em Lattice '87}, three talks gave birth to a
new field of research, the study of heavy quarks in lattice QCD.
Estia Eichten~\cite{Eic87} emphasized that lattice QCD should provide
reliable information on hadrons with heavy quarks, which would help
determine the Cabibbo-Kobayashi-Maskawa (CKM) matrix.
He suggested starting with the {\em static\/} approximation and adding
corrections in~$1/m_Q$.
One of the developments to grow from this suggestion is the heavy-quark
effective theory (HQET), an enormous subject in its own right.
Peter Lepage~\cite{Lep87} introduced non-relativistic QCD (NRQCD)
as a tool for studying spectroscopy and matrix elements of systems
with one or more heavy quarks.
This effective theory also has enjoyed widespread application.
Finally, Luciano Maiani~\cite{Gav87} presented, among other things,
the first well-known calculation from lattice QCD of the leptonic
decay constant of the $B$~meson.
Maiani and his collaborators took Wilson's action for light quarks
and bravely applied it to the $b$~quark, even though their lattice's
ultraviolet cutoff was below the mass~$m_b$.

The idea that lattice QCD could play a role in interpreting future
experiments was exciting, and it attracted much attention.
In addition to the spectra of quarkonium and ``heavy-light'' hadrons,
considerable effort has been devoted to the hadronic matrix elements
for leptonic and semileptonic decays of heavy-light mesons
($B$, $B_s$, $D$, and~$D_s$) and neutral meson mixing.
The decay constant of the generic heavy-light pseudoscalar meson,
denoted~$f_P$, parameterizes the hadronic amplitude for leptonic decays.
It has received the most attention, especially~$f_B$.
It was expected to be the most straightforward of matrix elements
and, thus, a bellwether.
Unfortunately, a reliable calculation could not be done quickly.
Fortunately, the technical and conceptual difficulties have been largely
overcome, and the calculation of~$f_B$ in lattice QCD has grown up at last.

Computationally $f_B$ is indeed just as easy as~$f_\pi$,
but the interpretation of the results has not been obvious.
Consequently, the literature contains a wide range of estimates, some of
which should not be taken seriously.
For example, several early calculations in the static limit contained
too much contamination from radial excitations~$B$~\cite{Has92},
and the results are misleadingly high.

More recently, a fuller understanding of the interplay between the
effective theories and the lattice has helped to reduce the effects
of lattice artifacts.
The effective theories NRQCD and HQET are derived from QCD by lowering
the renormalization point, or cutoff,~$\mu$ until
$|\vek{p}|\ll\mu\approx m_Q$, where~$|\vek{p}|$ is the heavy quark's
typical momentum and $m_Q$ its mass.
Because $|\vek{p}|/m_Q\ll1$ the interactions in the effective Lagrangian
can be organized in powers of~$|\vek{p}|/m_Q$.
(In quarkonium $|\vek{p}|\sim\alpha_s m_Q$; in heavy-light systems
$|\vek{p}|\sim\LQCD\sim200$~MeV.)
The two effective theories share the same Lagrangian, although the
power of~$|\vek{p}|/m_Q$ assigned to operators of higher dimension can
differ.
One can take the effective theories' Lagrangian and introduce the
lattice as an ultraviolet regulator, choosing the lattice spacing~$a$
so that $m_Q\approx\mu\sim\pi/a$.
This lattice theory is often called lattice NRQCD~\cite{Lep92}.
There are discretization effects, which are the higher-dimension
operators multiplied by calculable coefficients.
Although some of these operators are new, many are the same as in
the~$1/m_Q$ expansion.
Consequently, physical $1/m_Q$ effects and artificial~$a$ effects have
become intertwined.
Lattice practitioners must disentangle them and remove the lattice
artifacts, at least to the desired accuracy.

Alternatively, one can start with an action derived for $m_qa\ll1$,
such as Wilson's, and ask what happens when one applies it for
$m_Qa\approx1$.
This, essentially, is the way of Ref.~\cite{Gav87}.
Since then, many experts have said (and still do, out of habit),
that a heavy quark cannot be put directly
on the lattice because $m_Qa\approx1$.
On the other hand, numerous calculations have been published with
$m_Qa\approx1$, so there must be more to the story.
Indeed, the lattice theory does not break down.
Instead, the lattice artifacts are again intertwined with the~$1/m_Q$
effects, as in lattice NRQCD.
The same operators appear, but the coefficients
are different, though still calculable.
For a class of actions based on the Wilson action, it has been shown,
to all orders in perturbation theory, that the coefficients remain
small, for all~$m_Qa$~\cite{KKM97}.

In preparing a brief review of a subject one is faced with the choice
between a catalog of all recent results or a synthesis of developments
over a longer period of time.
The proceedings of the Lattice conferences provide excellent examples
of the former~\cite{Ono97,Ali97,Dra98}.
By contrast, this paper gives a view of the (theoretical and
computational) progress, focusing on~$f_B$.
Owing to space limitations the material presented on the allied
subject of neutral meson mixing and on phenomenologically promising
form factors of semileptonic decays is brief.

\section*{Numerical Lattice Calculations}

When one thinks of lattice QCD, one usually thinks of large-scale
numerical calculations.
This approach computes the functional integrals of quantum field theory
by applying a Monte Carlo method with importance sampling.
Statistical errors arise here, and with more and more computer time
these errors can be made arbitrarily small.
Over the years various clever techniques have been devised to enhance
the ``signal-to-noise'' ratio, that is, to reduce the statistical
error for fixed computing resources.

To use Monte Carlo methods, three modifications are introduced.
First, spacetime becomes a finite box, usually with periodic boundary
conditions.
Second, the spacetime continuum becomes a discrete lattice.
Last, the so-called quenched approximation is applied.
The first two are common to many kinds of numerical analysis, but
the last requires a short explanation.
The quenched approximation treats a hadron's valence quarks and
all the exchanged gluons fully, including retardation, but the
back-reaction of closed quark loops on the gluons is omitted.
For particle physics a more descriptive name (and one of the original
names) would be the valence approximation, but the term ``quenched,''
taken from statistical mechanics, is more commonly used.
The back-reaction of closed quark loops is computationally burdensome.
Because its omission saves computer time, the quenched approximation
is a useful way to control the other errors and, thus, to teach
theorists how to analyze uncertainties.

Let us return to the first two approximations.
A volume larger than a few fm on a side should be good enough.
After all, one does not expect the true size and boundary of the
universe to effect the physics of hadrons.
It is with the lattice itself that the subject becomes a craft.
The continuum limit can be reached by taking the lattice spacing
$a\to 0$ with brute force, or by improving the action to reduce
discretization effects, or by a combination of the two.
The crudest form of brute force, namely to take $m_Qa\ll1$, would
require, for the bottom quark, a fantastically small lattice spacing.
This has never been done.

\section*{$\vek{B}$, $\vek{D}$, $\vek{K}$ \& CKM}
Table~\ref{tab:CKM} contains a list of specific reactions with
conventional parameterizations of hadronic matrix elements.
Together each pair can determine the listed element of the CKM matrix.
\begin{table}
\caption{How to combine exclusive experimental measurements with
calculations in (lattice) QCD to obtain the Cabibbo-Kobayashi-Maskawa
matrix.}\label{tab:CKM}
\renewcommand{\arraystretch}{1.2}
\begin{tabular}{ccc|ccc}
\textcolor{green}{measure} &
\textcolor{brown}{compute} &
\textcolor{magenta}{determine} &
\textcolor{green}{measure} &
\textcolor{brown}{compute} &
\textcolor{magenta}{determine} \\ \hline
\textcolor{green}{$\pi\to\mu\nu$} &
\textcolor{brown}{$f_\pi$} &
\textcolor{magenta}{$|V_{ud}|$} & \\
\textcolor{green}{$K\to\mu\nu$} &
\textcolor{brown}{$f_K$} &
\textcolor{magenta}{$|V_{us}|$} &
\textcolor{green}{$K\to\pi e\nu$} &
\textcolor{brown}{$f_\pm^{K\pi}(q^2)$} &
\textcolor{magenta}{$|V_{us}|$} \\
\textcolor{green}{$B\to\tau\nu$} &
\textcolor{brown}{$f_B$} &
\textcolor{magenta}{$|V_{ub}|$} &
\textcolor{green}{$
\begin{array}{r@{\;\to\;}l} B & \pi e\nu \\ & \rho e\nu \end{array}
$ } &
\textcolor{brown}{$
\left.
\begin{array}{c} f_\pm^{B\pi}(q^2) \\ A_1^{B\rho}(q^2) \end{array} \right\}
$ } &
\textcolor{magenta}{$|V_{ub}|$} \\
\textcolor{green}{$D\to\mu\nu$} &
\textcolor{brown}{$f_D$} &
\textcolor{magenta}{$|V_{cd}|$} &
\textcolor{green}{$
\begin{array}{r@{\;\to\;}l} D & \pi e\nu \\ & \rho e\nu \end{array}
$ } &
\textcolor{brown}{$
\left.
\begin{array}{c} f_\pm^{D\pi}(q^2) \\ A_1^{D\rho}(q^2) \end{array} \right\}
$ } &
\textcolor{magenta}{$|V_{cd}|$} \\
\textcolor{green}{$D_s\to\mu\nu$} &
\textcolor{brown}{$f_{D_s}$} &
\textcolor{magenta}{$|V_{cs}|$} &
\textcolor{green}{$
\begin{array}{r@{\;\to\;}l} D & K e\nu \\ & K^* e\nu \end{array}
$ } &
\textcolor{brown}{$
\left.
\begin{array}{c} f_\pm^{DK}(q^2) \\ A_1^{DK^*}(q^2) \end{array} \right\}
$ } &
\textcolor{magenta}{$|V_{cs}|$} \\
\textcolor{green}{$B_c\to\mu\nu$} &
\textcolor{brown}{$f_{B_c}$} &
\textcolor{magenta}{$|V_{cb}|$} &
\textcolor{green}{$
\begin{array}{r@{\;\to\;}l} B & D e\nu \\ & D^* e\nu \end{array}
$ } &
\textcolor{brown}{$
\left.
\begin{array}{c} h_\pm(w) \\ h_{A_1}(w) \end{array} \right\}
$ } &
\textcolor{magenta}{$|V_{cb}|$} \\
\hline
\multicolumn{6}{c}{\bf Neutral meson mixing} \\
\textcolor{green}{$K^0 \leftrightarrow \bar{K}^0$} &
\textcolor{brown}{$\frac{8}{3}m^2_{K}f^2_{K}B_{K}$} &
\textcolor{magenta}{$\varepsilon_K(\rho,\,\eta)$} \\
\textcolor{green}{$B^0_d \leftrightarrow \bar{B}^0_d$} &
\textcolor{brown}{$\frac{8}{3}m^2_{B_d}f^2_{B_d}B_{B_d}$} &
\textcolor{magenta}{$|V_{td}|$} &
\textcolor{green}{$B^0_s \leftrightarrow \bar{B}^0_s$} &
\textcolor{brown}{$\frac{8}{3}m^2_{B_s}f^2_{B_s}B_{B_s}$} &
\textcolor{magenta}{$|V_{ts}|$} \\
\hline
\end{tabular}
\renewcommand{\arraystretch}{1.0}
\end{table}
Every element of the CKM matrix appears except~$V_{tb}$ and the entire
($3\times 3$ unitary) matrix can be constrained.

In the Standard Model, the leptonic partial width of a generic
pseudoscalar meson~$P$, containing quarks of flavors~$p$ and~$q$,
is given by
\begin{equation}
\Gamma_{P\to l\nu} = \left(
\begin{array}{c} {\rm known} \\ {\rm factors} \end{array} \right)
f_P^2 |V_{pq}|^2.
\end{equation}
If one could compute~$f_P$ (i.e., $f_K$ or $f_D$ or $f_B$) with
a reliably estimated uncertainty, a measurement of the partial width
is tantamount to a measurement of~$|V_{pq}|$.

Similarly, the differential decay rate of a semileptonic decay is given
by
\begin{equation}
\label{semiGamma}
\frac{d\Gamma_{P\to Hl\nu}}{dq^2} = \left(
\begin{array}{c} {\rm known} \\ {\rm factors} \end{array} \right)
{\cal F}^2(q^2) |V_{pq}|^2,
\end{equation}
where $q$ is momentum carried off by the leptons, and $\cal F$ is
the appropriate form factor.
These processes do not suffer the helicity suppression of the leptonic
decays, so the statistical error of the experimental measurements is
smaller.
Furthermore, lattice QCD can provide the~$q^2$ dependence, at least
when the momentum of the daughter hadron is not too large.
If theory and experiment exhibit the same shape as a function of~$q^2$,
one's confidence in the systematics increases qualitatively.

Neutral meson mixing reveals a glimpse of the third row of the CKM
matrix.
For example, the mass difference of neutral $B$ mesons is given by
\begin{equation}
\label{xB}
\textcolor{green}{x_q=\frac{\Delta M_{B^0_q}}{\Gamma_{B^0_q}}} = \left(
\begin{array}{c} {\rm known} \\ {\rm factors} \end{array} \right)
\textcolor{brown}{\frac{8}{3}m^2_{B_q}f^2_{B_q}B_{B_q}}
\textcolor{magenta}{|V_{tq}^*V_{tb}|^2} ,
\end{equation}
where the flavor $q$ can be either down or strange.
The notation employing $\frac{8}{3}m_{B_q}^2$, $f_{B_q}^2$, and the
``bag parameter''~$B_{B_q}$ is historical and is taken from the
kaon system.
(In the so-called vacuum saturation approximation, $B_B=1$.)
Nevertheless, this formula, more so than leptonic $B$ decay, motivates
the interest in~$f_{B_d}$ and~$f_{B_s}$.

\begin{figure} 
\centerline{\epsfig{file=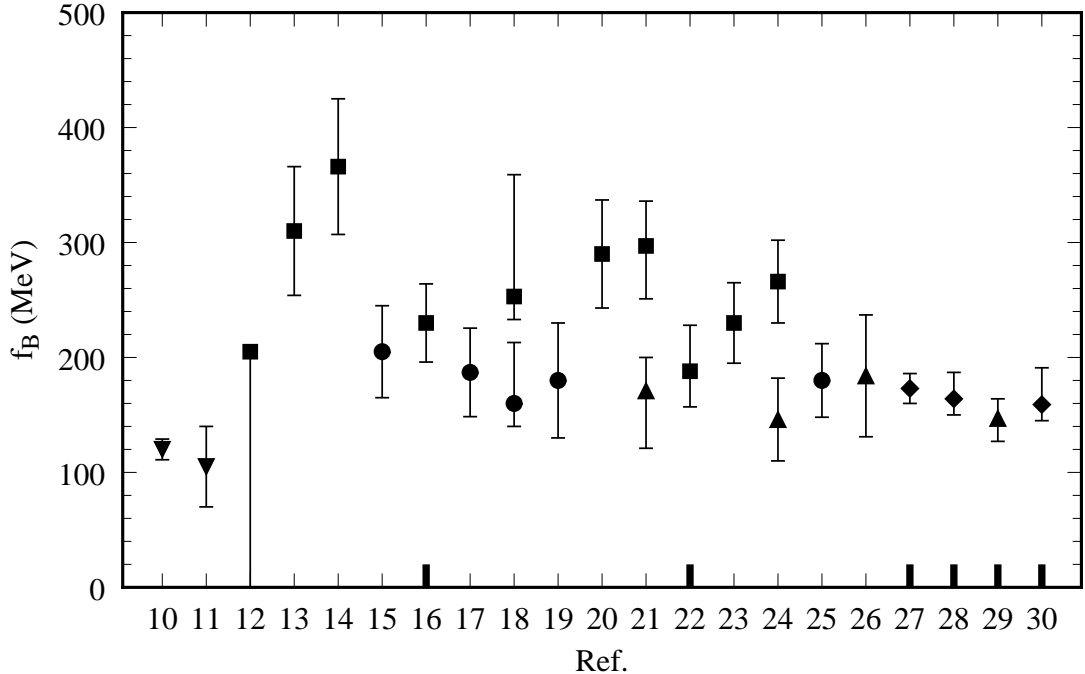,width=\textwidth}}
\vspace{10pt}
\caption[fig:fB]{Time-line of calculations of $f_B$ with lattice QCD.
Methods shown are
extrapolation from $m_Q\leq m_c$ to $m_b$ (inverted triangles),
the static limit $m_Q\to\infty$ (squares),
interpolation between $m_c$ and $\infty$ (circles),
NRQCD (triangles), and that of Ref.~\cite{KKM97}.
Entries with a thick tick-mark on the horizontal axis control lattice
spacing effects, as explained in the text.}\label{fig:fB}
\end{figure}

Figure~\ref{fig:fB} shows a time-line of calculations of~$f_B$ with
lattice QCD~%
\cite{Gav88,Ber88,Bou89,All90,Wup90,Aba91,Wup92,Ber93,UKQ93,Wup93,%
APE94,Has94,Dun94,All94,Dav96,All97,Ish97,Aok97,Kha97,Ali98,Ber98}.
The results included in Fig.~\ref{fig:fB} have been selected with two
criteria:
Conference proceedings, which are almost always followed (eventually)
by papers in refereed journals, have been omitted.
Otherwise, I have taken papers whose authors were self-confident enough
to quote a result in the abstract.
The second criterion is not necessarily fair to cautious innovators,
but, on the other hand, it generates the picture that is seen by
outsiders.
Figure~\ref{fig:fB} does contain a few exceptions to the second
criterion, to include calculations that offered new technical
developments.

One can divide the time-line into
infancy~\cite{Gav88,Ber88,Bou89,All90,Wup90},
childhood~\cite{Aba91,Wup92,Ber93,UKQ93,Wup93,APE94,Has94,Dun94},
youth~\cite{All94,Dav96,All97,Ish97}, and
adulthood~\cite{Aok97,Kha97,Ali98,Ber98}.
(A similar, more discriminating classification has been made by
Bernard~\cite{Ber97}.)
In adulthood, with refereed publications starting in November~1997,
the scatter that characterizes the field through its youth has settled
down.
The results with a thick tick-mark on the horizontal axis use several
lattice spacings~\cite{Wup92,Dun94,Aok97,Kha97,Ber98} or a fully
consistent implementation of lattice NRQCD~\cite{Ali98}.
Thus, one could say that Refs.~\cite{Wup92,Dun94} are mature results of
the static approximation but, because the contribution of order~$1/m_b$
is not negligible, not of~$f_B$.

\section*{Heavy Quarks and Lattice Field Theory}
In the introduction, I explained that physical~$1/m_Q$ effects and
artificial~$a$ effects are intertwined and that a better appreciation
of the intertwining was required before calculations of~$f_B$
could mature.
In particular, a theoretical analysis~\cite{KKM97} of Wilson quarks
away from the small-mass limit was needed to obtain the results in
Refs.~\cite{Aok97,Kha97,Ber98}.
This section summarizes some of the main ideas by comparing and
contrasting the effective theories in the continuum and on the lattice.

The effective Hamiltonian of QCD for heavy quarks can be written
\begin{equation}
\label{H}
H = M_1 \bar{\Psi}\Psi + \bar{\Psi}\gamma_0 A_0\Psi + \bar{\Psi}h\Psi,
\end{equation}
where~$\Psi$ is the fermion field and~$h$ is given by an expansion
in $1/m_Q$, namely
\begin{equation}
\label{h}
h = - \frac{\vek{D}^2}{2M_2}
  - z_B \frac{i\vek{\Sigma}\cdot\vek{B}}{2M_2} 
  + z_{\rm s.o.}
  \frac{\{\vek{\gamma}\cdot\vek{D},\vek{\alpha}\cdot\vek{E}\}}{8M_2^2}
  - z_4 \frac{(\vek{D^2})^2}{8M_2^3} + \cdots,
\end{equation}
corresponding to the kinetic energy, hyperfine splitting, spin-orbit
splitting, relativistic corrections, etc.%
\footnote{Analogous expansions are introduced for the operators
mediating electroweak transitions.}
This result follows from a series of Foldy-Wouthuysen-Tani
transformations, or from noticing the heavy-quark symmetries of QCD,
writing down allowed operators with arbitrary coefficients, and
matching physical observables to standard QCD.
With radiative corrections, one finds $z=1+O(g^2)$.

The rest mass of a quark is~$M_1$ and the kinetic energy of a quark
is~$\vek{p}^2/2M_2$.
It is convenient to call $M_2$ the kinetic mass, even though
Lorentz invariance implies $M_2=M_1$.
Let us write $H=M_1\bar{\Psi}\Psi+H_Q$.
The rest-mass term and~$H_Q$ commute, even in the interacting theory,
so eigenstates of~$H$ are simultaneously eigenstates of~$H_Q$.
Thus, one can drop the rest-mass term or readjust~$M_1$ according
to convenience, without changing the dynamics of heavy-quark~QCD.
The physically relevant parameter is the kinetic mass, which one adjusts
so that $M_2=m_Q$.

The lattice Hamiltonian takes the same structure as in Eqs.~(\ref{H})
and~(\ref{h}), but with changes to the operators' coefficients.
One can express this by replacing~$h$ in Eq.~(\ref{H}) with
$h_{\rm lat}=h+\delta h$ and writing
\begin{equation}
\label{hlat}
\delta h  = a b_B i\vek{\Sigma}\cdot\vek{B}
  + a^2b_{\rm s.o.} \{\vek{\gamma}\cdot\vek{D},\vek{\alpha}\cdot\vek{E}\}
  + a^3b_4(\vek{D^2})^2 + a^3w_4\sum_iD_i^4 + \cdots,
\end{equation}
where the coefficients $b=b(m_Qa,g^2)$ and $w=w(m_Qa,g^2)$ are functions
of the (lattice) quark mass~$m_Qa$ and the gauge coupling~$g^2$.
The same operators appear, as well as others, such as the last one,
that break rotational symmetry.
Part of the craft of the numerical work is to adjust the underlying
lattice action so that these artifact coefficients $b$ and $w$ vanish,
at least to some accuracy.

In lattice NRQCD and HQET this pattern arises by construction, in
particular one sets $M_1=0$.
It is fairly straightforward to adjust the~$b$s and~$w$s to vanish at
the tree level of perturbation theory in~$g^2$.
Beyond the tree level it is still possible, but arduous perturbative 
calculations are needed.
Power-law divergences appear in loop diagrams, so one cannot take
$a\to 0$ by brute force~\cite{Lep87}.
That means that the lattice artifacts can be removed only by further
refinements of the NRQCD action.
For heavy-light mesons, the action of the light quarks must be improved
to a consistent level~\cite{Mor98}, as in Ref.~\cite{Ali98}.

With actions for Wilson quarks, such as the clover action%
\footnote{The clover action adds a term to the Wilson action to
eliminate the leading lattice artifact~\cite{She85}.}
or the Wilson action itself, the pattern sketched here is less
immediate.
It is guaranteed, however, by the heavy-quark symmetry of the
lattice action.
Because the lattice violates Lorentz invariance, the rest and kinetic
masses differ; in practice $M_1<M_2$.
The $b$s and~$w$s are, however, \emph{bounded\/} for all masses.
As $m_Qa\to 0$, these coefficient functions go to a constant%
\footnote{If the limit is a constant, the lattice artifact still
vanishes, owing to the explicit powers of~$a$.}
or vanish like $(m_Qa)^{s_0}$, for some integer~$s_0$, by Symanzik's
standard analysis of cutoff effects.
On the other hand, as $m_Qa\to\infty$, they vanish like
$(1/m_Qa)^{s_\infty}$, for some~$s_\infty$, by heavy-quark
symmetry~\cite{KKM97}.
The full functional dependence on~$m_Qa$ is \emph{calculable\/} in
perturbation theory and, sometimes, nonperturbatively.
The coefficient functions $b$s and~$w$s depend on the details of the
lattice action.
For example, with the Wilson action $b_B\neq0$, but with the clover
action one can adjust an unphysical coupling until $b_B$~vanishes.

\section*{Leptonic Decays}

Let us now return to Fig.~\ref{fig:fB}.
The plotting symbols distinguish methods.
Squares denote calculations in the static approximation,
and triangles denote calculations with lattice NRQCD.
Inverted triangles~\cite{Gav88,Ber88}, circles, and diamonds denote
treatments of numerical data from the Wilson or clover action.

Because the latter connects naturally with the effective theories,
Ref.~\cite{KKM97} suggested treating the bottom quark on the lattice
by adjusting the bare mass until $M_2=m_b$.
In addition, a suitably normalized operator, essentially one built from
the heavy-quark field~$\Psi$, must be used for the current.
Results with this approach are denoted with diamonds.
The discretization errors are of order~$\LQCD a$ to some power, from
matrix elements of the operators in Eq.~(\ref{hlat}).
(These effects are then multiplied by a coefficient~$b$ or~$w$, which is
a number of order~1 or, in limiting cases, smaller still.)
With this interpretation of the numerical data from the clover action,
the lattice results for~$f_B$ are nearly independent of the lattice
spacing~\cite{Aok97,Kha97}.

In earlier work with Wilson quarks, the tuning of the mass and the
normalization of the current introduced discretization effects of
order $m_Qa$ or $(m_Qa)^2$.
The authors minimized these lattice artifacts by reducing the quark
mass of the lattice calculations below that of the charmed quark.
This is still large: $m_ca\sim5\LQCD a$.
Then the results were extrapolated up to~$m_b$ with fits to $1/m_Q$
expansions, either with (circles) or without (inverted triangles)
the help of the static value.
This intertwines lattice artifacts and $1/m_Q$ effects in ways that
depend on details of the fits, not least because one cannot verify
whether the $1/m_Q$ expansion converges for the low quark masses,
on which the fits are based.

Table~\ref{tab:fBfD} tabulates the mature
results~\cite{Aok97,Kha97,Ali98,Ber98} for~$f_B$, $f_{B_s}$, $f_D$,
and~$f_{D_s}$, along with an average~\cite{Par98} of results from
experimental measurements of $\Gamma_{D_s\to\mu\nu}$, combined
with~$|V_{cs}|$.
\begin{table}
\caption[tab:fBfD]{Compendium of recent results for decay constants of
heavy-light mesons.
The experimental average comes, in fact, from measurements of
$|V_{cs}|f_{D_s}$, which then take $|V_{cs}|$ from unitarity or from
neutrino production of charm}\label{tab:fBfD}
\begin{tabular}{cc|cccc|cccc}
 &   & \multicolumn{4}{c|}{$f_B$} & \multicolumn{4}{c}{$f_{B_s}$} \\
method & Ref. & MeV & stat & syst & quench & MeV & stat & syst & quench \\
\hline
\cite{KKM97} &
\cite{Aok97} &	173 & 04 & 09 & 09 &
		199 & 03 & 10 & 10 \\
\cite{KKM97} &
\cite{Kha97} &	164 & $^{+14}_{-11}$ & 08 & $^{+10}_{-00}\%$ &
		185 & $^{+13}_{-08}$ & 09 & $^{+10}_{-00}\%$ \\
NRQCD &
\cite{Ali98} &	147 & 11 & 11 & $^{+08}_{-12}$ &
		175 & 08 & 13 & 10 \\
\cite{KKM97} &
\cite{Ber98} &	159 & 11 & $^{+22}_{-09}$ & $^{+21}_{-00}$ &
		175 & 10 & $^{+28}_{-10}$ & $^{+25}_{-01}$ \\
 &   & \multicolumn{4}{c|}{$f_D$} & \multicolumn{4}{c}{$f_{D_s}$} \\
method & Ref. & MeV & stat & syst & quench & MeV & stat & syst & quench \\
\hline
\cite{KKM97} &
\cite{Aok97} &	197 & 02 & 14 & 10 &
		224 & 02 & 16 & 12 \\
\cite{KKM97} &
\cite{Kha97} &	194 & $^{+14}_{-10}$ & 10 & $^{+10}_{-00}\%$ &
		213 & $^{+14}_{-11}$ & 11 & $^{+10}_{-00}\%$ \\
\cite{KKM97} &
\cite{Ber98} &	195 & 11 & $^{+15}_{-08}$ & $^{+15}_{-00}$ &
		213 & 09 & $^{+23}_{-09}$ & $^{+17}_{-00}$ \\
  experiment &
\cite{Par98} &	    &    &                &                &
		243 & \multicolumn{3}{c}{\rm 36~(total)} \\
\hline
\end{tabular}
\end{table}
While it is tempting to average the results, it is subtle to do so
properly, because the systematic errors are largely, but not entirely,
common.
If you must have an average, consult Draper's review~\cite{Dra98},
or use your own eye.

The central values in Table~\ref{tab:fBfD} (except the experiment!)\
are in the quenched approximation.
The \emph{error bars}, however, reflect estimates of the associated
uncertainty.
The best estimation is that of the MILC Collaboration~\cite{Ber98} who
have some lattice results including up and down quark loops.
(The strange and heavier quarks are still quenched.)
The partially unquenched data sometimes lie higher than the quenched
data, sometimes not, leading to the very asymmetric error estimate.
These results are encouraging, not least because they suggest that
the wait for a fully unquenched calculation will not be too much longer.

\section*{Neutral Meson Mixing}
Because the calculation is technically more demanding, the literature
contains fewer calculations of the bag parameters of $B^0_d$ and
$B^0_s$~meson mixing than of the decay constants.
Recent publications report $B_{B_s}/B_{B_d}\approx 1$ and
\begin{equation}
\hat{B}_{B_d} = \left\{
    \begin{array}{ll}
	1.03 \pm 0.06 \pm 0.18 & \cite{Gim97} \\
	1.40 \pm 0.06\;^{+0.04}_{-0.26} & \cite{Chr97} \\
	1.23 \pm 0.05 \pm 0.15 & \cite{Chr97,Gim97} \\
	1.17 \pm 0.09 \pm 0.05 & \cite{Yam98}
    \end{array} \right.,
\end{equation}
where $\hat{B}_{B_d}$ is the renormalization-scheme independent
combination.
The first three entries use the static approximation.
The third entry comes from an analysis in Ref.~\cite{Chr97} of the
data in Ref.~\cite{Gim97}.
The last entry uses lattice NRQCD and finds, additionally, that the
dependence on $1/m_Q$ is not large.
See Ref.~\cite{Dra98} for more results.
There are also results for the ratio of matrix elements:
\begin{equation}
\frac{\frac{8}{3}m_{B_s}^2f_{B_s}^2B_{B_s}}%
{\frac{8}{3}m_{B_d}^2f_{B_d}^2B_{B_d}} = \left\{
    \begin{array}{ll}
	1.38 \pm 0.07 & \cite{Gim97} \\
	1.76 \pm 0.10\;^{+57}_{-42} & \cite{BBS98}
    \end{array} \right..
\end{equation}
On the whole, the impression is that more work needs to be done to gain
control over the systematic errors;
calculations of~$B_B$ are not as mature as those of~$f_B$.

The kaon's bag parameter $B_K$ is needed to predict~$\varepsilon_K$,
a measure of indirect CP violation in $K\to\pi\pi$.
Given $B_K$ from (lattice) QCD, a measurement of~$\varepsilon_K$ traces
a hyperbola in the complex~$V_{td}$ plane.
The history of $B_K$ has similarities to that of~$f_B$: numerical work
was a greater challenge than initially hoped, and theoretical insight
is needed as a guide.
For example, the lattice-spacing dependence (with staggered fermions)
is surprisingly steep, but one now knows to extrapolate to the
continuum limit in~$a^2$ (rather than $a$)~\cite{Sha94}.
Two recent calculations find
\begin{equation}
B_K({\rm NDR,~2~GeV}) = \left\{
    \begin{array}{l@{\,\pm\,}ll}
        0.62  & 0.02\pm 0.02 & \cite{Kil98} \\
	0.628 & 0.042        & \cite{Aok98}
    \end{array} \right.,
\end{equation}
where the main uncertainty comes from the continuum extrapolation.
Quenching and degenerate quark-mass effects may each lead to
underestimates of order~5\%~\cite{Sha97}.

\section*{Semileptonic Decays}
Semileptonic decays are wonderful for learning
about the first two rows of the CKM matrix.
Because a hadron is in the final state, lattice calculations of the
form factors are more difficult than decay constants, but not much
more difficult.

In these decays there is an additional kinematic variable, the
momentum~$|\vek{p}'|$ of the daughter hadron (in the parent's rest
frame).
The Lorentz invariant~$q^2$ is linearly related to~$|\vek{p}'|$.
Until recently, calculations of these form factors were done with
$m_Q\lsim m_c$ and extrapolated up to $m_c$ or $m_b$ with $1/m_Q$
expansions.
Since the kinematically allowed range of~$q^2$ depends on the
heavy-quark mass, another extrapolation is made.
It is clear that the extrapolations once again intertwine artificial~$a$
effects with physical $1/m_Q$ and $q^2$ dependence.
The details of the intertwining are not transparent (to me, anyway),
so I reserve comment and direct the reader to reviews by
Onogi~\cite{Ono97} and by Draper~\cite{Dra98}.

More recently, calculations have been done with lattice
NRQCD or with Wilson quarks interpreted~\cite{KKM97} as suggested by
Eqs.~(\ref{H})--(\ref{hlat}).
These techniques are especially powerful here, because the ability to
compute directly at the $B$ mass decouples extrapolations in $q^2$
from the mass dependence.
Calculations are underway for form factors of heavy-to-light
transitions, such as $D\to \pi l\nu$ and $D\to Kl\nu$, which together
yield $|V_{cd}/V_{cs}|$~\cite{Tom97,Sim98}, and $B\to\pi l\nu$,
which yields $|V_{ub}|$~\cite{Tom97,Mat97,Rya98}.

I would like to conclude with a preliminary result~\cite{Has98} on
the zero-recoil form factor for the decay $B\to Dl\nu$, which will
improve the determination of~$|V_{cb}|$.
A similar study of $B\to D^{*}l\nu$ is in progress.
Until now there have been calculations of the shape of the form factors
for $B\to D^{(*)}l\nu$, but the normalization has been computed only
poorly, in perturbation theory.
It is possible, however, to handle almost all
of the normalization nonperturbatively.
For example~\cite{Man93},
\begin{equation}
\frac{\langle D|{\cal V}_0|B \rangle \langle B|{\cal V}_0|D \rangle}%
{\langle D|{\cal V}_0|D \rangle \langle B|{\cal V}_0|B \rangle} =
\frac{h_+^{B\to D}(1) h_+^{D\to B}(1)}{h_+^{D\to D}(1) h_+^{B\to B}(1)}
= |h_+^{B\to D}(1)|^2,
\end{equation}
and here at Fermilab we have found analogous ratios for
$h_-^{B\to D}(1)$ and~$h_{A_1}^{B\to D^*}(1)$.
For the class of actions considered in Ref.~\cite{KKM97}, the
remaining radiative corrections have been computed~\cite{Kro98},
and they are small.

In Eq.~(\ref{semiGamma}) for $B\to Dl\nu$ one requires
${\cal F} = h_+ - (m_B-m_D)h_-/(m_B+m_D)$.
The advantage of the new method is that, in effect, it calculates not
${\cal F}(1)$ but the deviation of ${\cal F}(1)$ from 1.
We find
\begin{equation}
{\cal F}(1) = 1.069\pm0.008\pm0.002\pm0.025,
\end{equation}
where the uncertainties are Monte Carlo statistics, tuning of the
quark masses, and a certain parametrically small contribution omitted
from~$h_-$.
Uncertainties from lattice artifacts and from quenching have not yet
been taken into account.

\end{document}

%% file: MyBWMap.ltx
\definecolor{red}{gray}{0}
\definecolor{green}{gray}{0}
\definecolor{blue}{gray}{0}
\definecolor{cyan}{gray}{0}
\definecolor{magenta}{gray}{0}
\definecolor{yellow}{gray}{0}
\definecolor{brown}{gray}{0}
\definecolor{gray}{gray}{0}
\definecolor{orange}{gray}{0}
\definecolor{pink}{gray}{0}
\definecolor{purple}{gray}{0}
\definecolor{clear}{gray}{1}